\documentclass[preprint,showpacs,preprintnumbers,nobibnotes]{revtex4}
\usepackage{amsmath}
\usepackage{graphicx}
\usepackage{dcolumn}
\usepackage{bm}

\input{tcilatex}
\begin{document}

\title{Phase diagrams, quantum correlations and critical phenomena of
antiferromagnetic Heisenberg model on diamond-type hierarchical lattices}
\author{Pan-Pan Zhang$^{1}$, Zhong-Yang Gao$^{2}$, Yu-Liang Xu$^{1}$,
Chun-Yang Wang$^{1}$}
\author{Xiang-Mu Kong$^{1,2}$}
\altaffiliation{Corresponding author. E-mail address: kongxm668@163.com (X.-M. Kong).}
\affiliation{$^{1}$School of Physics and Optoelectronic Engineering, Ludong University,
Yantai 264025, China\\
$^{2}$College of physics and Engineering, Qufu Normal University, Qufu
273165, China}
\date{\today }

\begin{abstract}
The spin-$1/2$ antiferromagnetic Heisenberg systems are studied on three
typical diamond-type hierarchical lattices~(systems A, B and C) with fractal
dimensions $1.63,$ $2$ and $2.58$, respectively, and the phase diagrams,
critical phenomena and quantum correlations are calculated by a combination
of the equivalent transformation and real-space renormalization group
methods. We find that there exist a reentrant behavior for system A and a
finite temperature transition in the isotropic Heisenberg limit for system C
(not for system B). Unlike the ferromagnetic case, the N\'{e}el temperatures
of antiferromagnetic\ systems A and B are inversely proportional to $\ln
\left( \Delta _{\text{c}}-\Delta \right) $ (when $\Delta \rightarrow \Delta
_{\text{c}}$) and $\ln \Delta $ (when $\Delta \rightarrow 0$),\
respectively. And we also find that there is a turning point of quantum
correlation in the isotropic Heisenberg limit $\Delta =0$ where there is a
"peak" of the contour and no matter how large the size of system is, quantum
correlation will change to zero in the Ising limit for the three systems.
The quantum correlation decreases with the increase of lattice size $L$ and
it is almost zero when $L\geq 30$ for system A, and for systems B and C,
they still exist when $L$ is larger than that of system A. Moreover, as an
example, we discuss the error of result in system A, which is induced by the
noncommutativity.
\end{abstract}

\keywords{Phase diagram; critical phenomena; quantum correlation;
antiferromagnetic Heisenberg model; diamond-type hierarchical lattices;
renormalization group }
\maketitle

\section{Introduction}

Fractals, due to their unique self-similar structure, have been widely
studied theoretically and experimentally in physics and materials science
over the past few decades, especially in the field of phase transitions and
critical phenomena of spin systems\cite{Z.R.Yang1996,nature2018}. For
classical spin systems on fractals, in particular, using the renormalization
group (RG) method to study critical phenomena, there have been many
conclusive results\cite{Z.R.Yang1996,gefen1980,gefen1983}. For quantum
ferromagnetic spin systems on fractals, there are also some studies and
conclusions, e.g., the research shows that the critical temperature $T_{%
\text{c}}=0\ $and $T_{\text{c}}>0$\ for the isotropic ferromagnetic
Heisenberg system\ on diamond-type hierarchical~(DH) lattice with fractal\
dimensions $d_{\text{f}}=2$ and $d_{\text{f}}=3$, respectively\cite%
{plascak1984,de sousa2003,mariz}.

Relatively speaking, the caculation of antiferromagnetic (AF) system is more
difficult than the ferromagnetic one. In recent years, various
renormalization group approaches\cite{DMRG,de Sousa,de Sousa-MRG,de
Sousa-MRG2}, Green's function technique\cite{spin system-1,Green1},
spin-wave theory~\cite{spin-wave} and Monte Carlo method \cite%
{Cuccoli2003,Krokhmalskii2017} have been applied to study the quantum AF
Heisenberg systems on translationally symmetric lattices. Sousa et al.
calculated the phase diagrams of the AF Heisenberg model by the mean-field
RG approach and found that the AF N\'{e}el temperature $T_{\text{N}}<T_{%
\text{c}}$~and $T_{\text{N}}>T_{\text{c}}$ for two-dimensional (2D) and
three-dimensional (3D) lattices, respectively\cite{de Sousa-MFRG2,de
Sousa-MRG2,deSousa2000}. However, results from spin-wave theory~and Green's
function technique obtained $T_{\text{N}}=T_{\text{c}}$~for AF Heisenberg
systems on 2D and 3D lattices \cite{spin-wave,Green1,Green2}. Quantum Monto
Carlo results indicate that an ordered low temperature phase is obtained
even at much smaller anisotropies $\Delta \sim 10^{-3}$~for AF Heisenberg
systems on 2D lattices \cite{Cuccoli2003,Cuccoli2003-2}. But there have been
no reports of the study on AF spin systems on fractal lattices so far.

On the other hand, as an important quantum resource, quantum correlation has
been applied to quantum communication, quantum teleportation and quantum
computation, etc\cite{1993,2020}. Quantum correlation contains more quantum
information than quantum entanglement which is a particular quantum
correlation\cite{Szasz2019,Jaszek2020}. The quantum discord (QD) which
captures nonclassical correlation even without entanglement is proved as an
effective measure for all aspects of quantum correlation\cite%
{Ollivier2001,Henderson2001}. As a wider class of measures than
entanglement, QD has become an active research topic over the past few years%
\cite{Modi2012,XuyuliangQD,Adesso2016,huang2014}, which has important
application in quantum information processing, quantum dynamics and even
biophysics\cite{Bran2015,Madhok2013,Girolami2014,Bra2010}. But, it isn't
difficult to find that the present studies have focused on the
one-dimensional spin chain, the studies of fractal lattices are less\cite%
{XuyuliangQD,usman2020}. It is significant to study the quantum correlation
of fractal lattice systems, which is helpful to understand the influence of
fractal structure on quantum information.

In this work, we investigate the phase diagrams, critical phenomena and QDs
of the AF Heisenberg model on three kinds of DH lattices~(systems A, B and
C) with fractal dimensions $d_{\text{f}}=1.63,$ $2$ and $2.58$,
respectively, which is shown in Fig.~$1$. Using the real-space RG method
straightforwardly, we study system A and obtain the relations between
quantum correlation with temperature and the anisotropic parameter. For
systems B and C, we apply a combination of the equivalent transformation and
real-space RG methods. On this basis, we hope to explore the relatively
universal laws through the further study of the fractal lattice systems.

The structure of this manuscript is organized as follows. In Sec.~\ref{sec2}
we summarize the method and study the phase diagram\ and critical behavior
of system A. In Sec.~\ref{sec3} the method and the results are presented for
systems B and C. Sec.~\ref{sec4} discusses the quantum correlations of the
three systems. In Sec.~\ref{sec5} the quantum effect and error of result are
discussed. The summary is given in the last section.

\section{The method and critical behaviors of system A\label{sec2}}

The spin-$1/2$ Heisenberg model is described by the effective Hamiltonian%
\begin{equation}
H=K\underset{\langle ij\rangle }{\sum }\left[ (1-\Delta )(\sigma
_{i}^{x}\sigma _{j}^{x}+\sigma _{i}^{y}\sigma _{j}^{y})+\sigma
_{i}^{z}\sigma _{j}^{z}\right] ,  \label{equ1}
\end{equation}%
where $K=J/k_{\text{B}}T$, $J$ is the exchange coupling constant ( $J>0$~and 
$J<0$ correspond to ferromagnetic model and AF one, respectively), $k_{\text{%
B}}$ the Boltzmann constant, $T$ the temperature, $\langle ij\rangle $
denotes nearest-neighbor sites of this system, $\Delta $ is the anisotropic
parameter, and $\sigma _{i}^{\alpha }$~$\left( \alpha =x,y,z\right) $ are
Pauli operators at site $i$. Note that, as particular cases, the Hamiltonian
describes the Ising ($\Delta =1$), isotropic Heisenberg ($\Delta =0$) and XY
($\Delta =-\infty $) models, respectively.

In this section, we study the phase diagram and critical behavior of system
A by the real-space RG method\cite{important 1983,important 1985}. Such kind
of lattice is constructed in an iterative manner which can be realized by
continuously replacing the dimer (a two-sites) with the generator\cite%
{Z.R.Yang1991}~(Fig.~$1$). The RG transformation process system A is shown
in Fig.~$2(a)$. As we can see, after summation of the internal spins, the
generator~(Fig. $2(a1)$) is transformed into a new structure --- a dimer
(Fig. $2(a2)$), which contains $\sigma _{1}$ and $\sigma _{2}$. This
procedure can be described as%
\begin{equation}
\exp (H_{12}^{\prime })=\underset{3456}{\text{Tr}}\exp (H_{123456}),
\label{equ2}
\end{equation}%
where%
\begin{equation}
H_{12}^{\prime }=K^{\prime }[(1-\Delta ^{\prime })(\sigma _{1}^{x}\sigma
_{2}^{x}+\sigma _{1}^{y}\sigma _{2}^{y})+\sigma _{1}^{z}\sigma
_{2}^{z}]+K_{0}  \notag
\end{equation}%
and%
\begin{eqnarray}
H_{123456} &=&K(1-\Delta )[(\sigma _{1}^{x}\sigma _{3}^{x}+\sigma
_{1}^{y}\sigma _{3}^{y})+(\sigma _{1}^{x}\sigma _{6}^{x}+\sigma
_{1}^{y}\sigma _{6}^{y})+(\sigma _{3}^{x}\sigma _{4}^{x}+\sigma
_{3}^{y}\sigma _{4}^{y})  \notag \\
&&+(\sigma _{5}^{x}\sigma _{6}^{x}+\sigma _{5}^{y}\sigma _{6}^{y})+(\sigma
_{2}^{x}\sigma _{5}^{x}+\sigma _{2}^{y}\sigma _{5}^{y})+(\sigma
_{2}^{x}\sigma _{4}^{x}+\sigma _{2}^{y}\sigma _{4}^{y})]  \notag \\
&&+K(\sigma _{1}^{z}\sigma _{3}^{z}+\sigma _{1}^{z}\sigma _{6}^{z}+\sigma
_{3}^{z}\sigma _{4}^{z}+\sigma _{5}^{z}\sigma _{6}^{z}+\sigma _{2}^{z}\sigma
_{5}^{z}+\sigma _{2}^{z}\sigma _{4}^{z})  \notag
\end{eqnarray}%
are the Hamiltonians of the renormalized dimer and of the generator,
respectively. $\underset{3456}{\text{Tr}}$ denotes a partial trace over the
internal $\sigma _{i}(i=3,4,5,6)$. To make Eq.~(\ref{equ2}) possible, an
additional constant $K_{0}$ has been included in $H_{12}^{\prime }$. The RG
recursion relations between the renormalized~$(\Delta ^{\prime },K^{\prime
}) $ and the original~$(\Delta ,K)$\ parameters are determined by Eq.~(\ref%
{equ2}). Notice that the noncommutativity between the Hamiltonians of the
neighboring generators is neglected, therefore the results are
approximations which will be discussed in Sec.~\ref{sec5}.

In order to calculate the partial trace in Eq.~(\ref{equ2}), we expand $\exp
(H_{12}^{\prime })$ as 
\begin{equation}
\exp (H_{12}^{\prime })=a^{\prime }+b_{12}^{\prime }(\sigma _{1}^{x}\sigma
_{2}^{x}+\sigma _{1}^{y}\sigma _{2}^{y})+c_{12}^{\prime }\sigma
_{1}^{z}\sigma _{2}^{z},  \label{equ3}
\end{equation}%
where $a^{\prime }$, $b_{12}^{\prime }$ and $c_{12}^{\prime }$ are functions
of $\Delta ^{\prime }$, $K^{\prime }$ and $K_{0}$. In the direct product
representation of $\sigma _{1}^{z}$ and $\sigma _{2}^{z}$, both the
left-hand side and right-hand side of Eq.~(\ref{equ3}) can be expressed as $%
4\times 4$ matrixes and we finally obtain 
\begin{align}
\exp (4K^{\prime })& =\frac{\left( a^{\prime }+c_{12}^{\prime }\right) ^{2}}{%
\left( a^{\prime }-c_{12}^{\prime }\right) ^{2}-16b_{12}^{{\prime }\text{{}}%
2}},  \label{equ4} \\
\exp (4\Delta ^{\prime }K^{\prime })& =\frac{\left( a^{\prime
}+c_{12}^{\prime }\right) ^{2}}{\left( a^{\prime }+4b_{12}^{\prime
}-c_{12}^{\prime }\right) ^{2}},  \label{equ5} \\
\exp (K_{0})& =\frac{a^{\prime }+c_{12}^{\prime }}{\exp (K^{\prime })}.
\label{equ6}
\end{align}

Similarly, we expand $\exp \left( H_{123456}\right) $ as%
\begin{eqnarray}
\exp (H_{123456}) &=&a+\underset{\langle ij\rangle }{\sum }\left[
b_{ij}\left( \sigma _{i}^{x}\sigma _{j}^{x}+\sigma _{i}^{y}\sigma
_{j}^{y}\right) +c_{ij}\sigma _{i}^{z}\sigma _{j}^{z}\right]  \label{7} \\
&&+\underset{\langle ij\rangle \neq \langle kl\rangle }{\sum }A_{ij,kl}+%
\underset{\langle ij\rangle \neq \langle kl\rangle \neq \langle mn\rangle }{%
\sum }B_{ij,kl,mn}+r\sigma _{1}^{z}\sigma _{2}^{z}\sigma _{3}^{z}\sigma
_{4}^{z}\sigma _{5}^{z}\sigma _{6}^{z},  \notag
\end{eqnarray}%
in which%
\begin{equation}
A_{ij,kl}=e_{ij,kl}(\sigma _{i}^{x}\sigma _{j}^{x}+\sigma _{i}^{y}\sigma
_{j}^{y})(\sigma _{k}^{x}\sigma _{l}^{x}+\sigma _{k}^{y}\sigma
_{l}^{y})+d_{ij,kl}(\sigma _{i}^{x}\sigma _{j}^{x}+\sigma _{i}^{y}\sigma
_{j}^{y})\sigma _{k}^{z}\sigma _{l}^{z}+f_{ij,kl}\sigma _{i}^{z}\sigma
_{j}^{z}\sigma _{k}^{z}\sigma _{l}^{z}  \notag
\end{equation}%
and%
\begin{eqnarray}
B_{ij,kl,mn} &=&g_{ij,kl,mn}(\sigma _{i}^{x}\sigma _{j}^{x}+\sigma
_{i}^{y}\sigma _{j}^{y})\sigma _{k}^{z}\sigma _{l}^{z}\sigma _{m}^{z}\sigma
_{n}^{z}+p_{ij,kl,mn}(\sigma _{i}^{x}\sigma _{j}^{x}+\sigma _{i}^{y}\sigma
_{j}^{y})(\sigma _{k}^{x}\sigma _{l}^{x}+\sigma _{k}^{y}\sigma
_{l}^{y})\sigma _{m}^{z}\sigma _{n}^{z}  \notag \\
&&{}+q_{ij,kl,mn}(\sigma _{i}^{x}\sigma _{j}^{x}+\sigma _{i}^{y}\sigma
_{j}^{y})(\sigma _{k}^{x}\sigma _{l}^{x}+\sigma _{k}^{y}\sigma
_{l}^{y})(\sigma _{m}^{x}\sigma _{n}^{x}+\sigma _{m}^{y}\sigma _{n}^{y}), 
\notag
\end{eqnarray}%
where $a$, $\{b_{ij}\}$, $\{c_{ij}\}$, $\ldots $ $\left\{
q_{ij,kl,mn}\right\} $ and $r$ depend on $K$ and $\Delta $. From Eqs.~(\ref%
{equ2}), (\ref{equ3}) and (\ref{7}), it can be obtained that $a^{\prime
}=16a $, $b_{12}^{\prime }=16b_{12}$, and $c_{12}^{\prime }=16c_{12}$, and
the recursion relations become%
\begin{align}
\exp (4K^{\prime })& =\frac{\left( a+c_{12}\right) ^{2}}{\left(
a-c_{12}\right) ^{2}-4b_{12}^{\text{{}}2}},  \label{equ8} \\
\exp (4\Delta ^{\prime }K^{\prime })& =\frac{\left( a+c_{12}\right) ^{2}}{%
\left( a+2b_{12}-c_{12}\right) ^{2}}.  \label{equ9}
\end{align}

By numerically iterating Eqs. (\ref{equ8}) and (\ref{equ9}), the AF phase
diagram is obtained, which is shown in Fig.~$3$. There are two stable fixed
points at $\left( 1,\text{\thinspace }\infty \right) $ and $\left( 1,\text{%
\thinspace }0\right) $ and two unstable fixed points at $\left( 1,\text{%
\thinspace }0.94\right) $ and $\left( 0,\text{\thinspace }0\right) $, which
corresponding to the Ising and the isotropic Heisenberg models,
respectively. For comparison, the inset shows the critical line of the
ferromagnetic system. We can see that there exists two phases in AF system,
namely, the ordered and disordered phases, respectively. The Ising fixed
point of the ferromagnetic system is the same as that of the AF one. At the
Ising fixed point, we obtain the correlation length critical exponent%
\begin{equation}
\nu =\frac{\ln b}{\ln \lambda }=1.779,  \label{10}
\end{equation}%
where $b=3$ is the scaling factor and $\lambda \equiv \left( \partial
K^{\prime }/\partial K\right) \mid _{\Delta =1,k_{\text{B}}T/\left\vert
J\right\vert =0.94}$. At the isotropic Heisenberg fixed point of the
ferromagnetic model, we obtain $\nu =0$. For the AF model, the phase diagram
is different from the ferromagnetic one: the critical line goes to zero at $%
\Delta _{\text{c}}=0.703$ (quantum critical point). Moreover, we find a
reentrant behavior in the phase diagram. This fact is usually due to the
quantum fluctuation which is related to the critical temperature and the
anisotropy parameter and we will discuss it in Sec.~\ref{sec5}. Furthermore,
we also study the critical temperature of this system when $T\rightarrow 0$.
As shown in Fig.~$4$, for the ferromagnetic case the curie temperature
satisfies 
\begin{equation}
T_{\text{c}}\sim \Delta  \label{11}
\end{equation}%
and for the AF case the N\'{e}el temperature $T_{\text{N}}$ satisfies%
\begin{equation}
T_{\text{N}}\sim \frac{1}{\ln \left( \Delta _{\text{c}}-\Delta \right) },
\label{12}
\end{equation}%
where $\Delta _{\text{c}}=0.703$.

\section{The method and the results for systems B and C\label{sec3}}

In this section, we use an effective combination of the equivalent
transformation and real-space RG methods to study the AF Heisenberg model on
systems B and C, which has been used for classical systems in Ref.~\cite%
{Z.R.Yang1991} but for quantum systems this is done for the first time as
far as we know. As we can see, antiferromagnetic interactions~$\left(
K<0\right) $ change to ferromagnetic interactions~($K^{\prime }>0$) by the
ET is shown\ in Fig.~$2(b)$. Therefore, we can obtain the critical point of
an AF system if we know that of the equivalent system with ferromagnetic
couplings. This method contains two steps~(see Fig.~$2(b)$):~(1)~The ET
technique is used to transform the AF system into an equivalent
ferromagnetic one and the equivalent transformation equations are
obtained.~(2)~By numerically iterating the RG recursion equations of the
equivalent ferromagnetic system can be obtained.~The results obtained from
the second step are substituted into the ET equations in the first step and
the AF phase diagram can be obtained.

In the following, we give the calculation procedure of the AF Heisenberg
model for system B. To construct the equivalent transformation equations, we
transform the cluster of Fig.~$2(b1)$~(which contains four $\sigma
_{i}(i=1,2,3,4)$) into Fig.~$2(b2)$~(the cluster contains two spins 1 and
2). Following the same steps as in Sec.~\ref{sec2}, we can obtain the ET
equations in the same form as Eqs.~(\ref{equ4})--(\ref{equ6}), as well as
the $16\times 16$ matrix%
\begin{eqnarray}
\exp \left( H_{1234}\right) &=&a+\underset{\langle ij\rangle }{\sum }\left[
b_{ij}(\sigma _{i}^{x}\sigma _{j}^{x}+\sigma _{i}^{y}\sigma
_{j}^{y})+c_{ij}\sigma _{i}^{z}\sigma _{j}^{z}\right]  \notag \\
&&{}+\underset{\langle ij\rangle \neq \langle kl\rangle }{\sum }%
d_{ij,kl}(\sigma _{i}^{x}\sigma _{j}^{x}+\sigma _{i}^{y}\sigma
_{j}^{y})\sigma _{k}^{z}\sigma _{l}^{z}  \label{13} \\
&&{}+\underset{\langle ij\rangle \neq \langle kl\rangle }{\sum }%
e_{ij,kl}(\sigma _{i}^{x}\sigma _{j}^{x}+\sigma _{i}^{y}\sigma
_{j}^{y})(\sigma _{k}^{x}\sigma _{l}^{x}+\sigma _{k}^{y}\sigma _{l}^{y}) 
\notag \\
&&{}+f\sigma _{1}^{z}\sigma _{2}^{z}\sigma _{3}^{z}\sigma _{4}^{z},  \notag
\end{eqnarray}%
where $a$, $\left\{ b_{ij}\right\} $, $\left\{ c_{ij}\right\} $, $\left\{
d_{ij,kl}\right\} $, $\left\{ e_{ij,kl}\right\} $ and $f$ are functions of $%
K $ and $\Delta $. Then using some similar calculations, it is obtained that 
$a^{\prime }=4a$, $b_{12}^{\prime }=4b_{12}$ and $c_{12}^{\prime }=4c_{12}$,
and we finally obtain the equivalent transformation equations as%
\begin{align}
\exp (4K^{\prime })& =\frac{\left( a+c_{12}\right) ^{2}}{\left(
a-c_{12}\right) ^{2}-4b_{12}^{\text{{}}2}},  \label{equ14} \\
\exp (4K^{\prime }\Delta ^{\prime })& =\frac{\left( a+c_{12}\right) ^{2}}{%
\left( a+2b_{12}-c_{12}\right) ^{2}},  \label{equ15}
\end{align}%
in which $a$, $b_{12}$, $c_{12}$ are functions of $K$ and $\Delta $ and the
equivalent system has ferromagnetic interactions $K^{\prime }>0$. For the
equivalent ferromagnetic system, we obtain the RG recursion relations in the
same form as Eqs.~(\ref{equ14}) and (\ref{equ15})%
\begin{align}
\exp (4K^{\prime \prime })& =\frac{\left( a+c_{12}\right) ^{2}}{\left(
a-c_{12}\right) ^{2}-4b_{12}^{\text{{}}2}},  \label{equ14-1} \\
\exp (4K^{\prime \prime }\Delta ^{\prime \prime })& =\frac{\left(
a+c_{12}\right) ^{2}}{\left( a+2b_{12}-c_{12}\right) ^{2}},  \label{equ15-1}
\end{align}%
where $a$, $b_{12}$, $c_{12}$ are functions of $K^{\prime }$ and $\Delta
^{\prime }$. Then, we substitute $K^{\prime }$ and $\Delta ^{\prime }$ which
are obtained by numerically iterating the ferromagnetic RG equations~(\ref%
{equ14-1}) and~(\ref{equ15-1}) into Eqs.~(\ref{equ14}) and (\ref{equ15}),
and finally get the phase diagram of the AF system.

Using the same procedure, we calculate the phase diagram of system C.
However, the analytical expressions of the right-hand side in Eqs.~(\ref%
{equ14}) and (\ref{equ15}) are difficult to calculate. In order to get the
expressions, we expend $\exp \left( H_{12345}\right) $ approximately as%
\begin{equation}
\exp \left( H_{12345}\right) \approx 1+H_{12345}+\frac{H_{12345}^{2}}{2!}%
+\cdots +\frac{H_{12345}^{11}}{11!},  \label{16}
\end{equation}%
in which we adopt the first twelve terms for convenient calculation.

Fig.~$5$ shows the phase diagrams of systems B and C in $\left( \Delta ,%
\text{\thinspace }k_{\text{B}}T/\left\vert J\right\vert \text{\thinspace }%
\right) $ space. The AF critical line of system A is also provided for
comparison. There are two unstable fixed points for system B at $\left( 1,%
\text{\thinspace }1.64\right) $ and $\left( 0,\text{\thinspace }0\right) $
(for system C at $\left( 1,\text{\thinspace }2.77\right) $ and $\left( 0,%
\text{\thinspace }2.06\right) $), corresponding to the Ising and the
isotropic Heisenberg fixed points, respectively. From Fig.~$5$, we find that:

(1) For low-dimensional system A, there exists a reentrant behavior in the
phase diagram.

(2) For system B, the N\'{e}el temperature tends to zero as decreasing $%
\Delta $.

(3) For higher-dimensional system C, there exists finite temperature phase
transition in the isotropic Heisenberg limit $\Delta =0$. Above studies are
in accordance with the previous quantum Monte Carlo\cite%
{Cuccoli2003,MC-2003-cite16,Clark1979,Cuccoli2003-2} and mean field
renormalization group\cite{de Sousa-MRG,de Sousa-MRG2,de Sousa-MFRG2,de
Sousa-MRG3} results. From our results, we can see that as the value of $d_{%
\text{f}}$ increases the phase transition will present in the isotropic
Heisenberg limit and as the value of $d_{\text{f}}$ decreases the reentrant
behavior will appear, and $d_{\text{f}}=2$ maybe the critical dimension. We
consider this phenomena may are suitable for different fractal dimensions
that is related to the quantum effects which will be discussed in Sec.~\ref%
{sec5}. Moreover, we also study the critical temperature when $\Delta
\rightarrow 0$ and obtain the following law for system B (Fig.~$6$):%
\begin{equation}
T_{\text{N}}\sim \frac{1}{\ln \Delta }.  \label{17}
\end{equation}

\section{The quantum correlations for the three lattices\label{sec4}}

In this section, we study the QD between two non-nearest-neighbor end spins
of AF Heisenberg model in three DH lattices at finite temperature and
discuss the relations of QD with temperature and anisotropic parameter. It
is widely accepted that quantum mutual information is a measure of the total
correlation contained in a composite quantum system which includes classical
and quantum correlations\cite{QMI1,QMI2}. We want to get quantum correlation
by subtracting classical correlation from quantum mutual information. As a
common representation of quantum correlation, QD captures nonclassical
correlation even without entanglement. For a quantum state $\rho _{\text{12}%
} $ of the composite system containing subsystems\ 1 and 2, the QD is
defined as 
\begin{equation*}
D\left( \rho _{\text{12}}\right) =I\left( \rho _{\text{12}}\right) -C\left(
\rho _{\text{12}}\right) ,
\end{equation*}%
where $I\left( \rho _{\text{12}}\right) $ is the quantum mutual information
and $C\left( \rho _{\text{12}}\right) $\ is the classical correlation\cite%
{QD,XuyuliangQD}. Because the density matrix $\rho _{\text{12}}$ of two
spins in AF Heisenberg model exhibits an X structure, it is called X state,
as%
\begin{equation*}
\rho _{\text{12}}=\left( 
\begin{array}{cccc}
a_{11} & 0 & 0 & a_{14} \\ 
0 & a_{22} & a_{23} & 0 \\ 
0 & a_{32} & a_{33} & 0 \\ 
a_{41} & 0 & 0 & a_{44}%
\end{array}%
\right) ,
\end{equation*}%
where%
\begin{eqnarray*}
a_{11} &=&a_{44}=\frac{e^{K}}{2e^{K}+e^{K}+e^{K\left( 1-2\Delta \right) }},
\\
a_{14} &=&a_{41}=0, \\
a_{22} &=&a_{33}=\frac{e^{K\left( 1-2\Delta \right) }+e^{K\left( 2\Delta
-3\right) }}{2\left( 2e^{K}+e^{K\left( 2\Delta -3\right) }+e^{K\left(
1-2\Delta \right) }\right) }, \\
a_{23} &=&a_{32}=\frac{e^{K\left( 1-2\Delta \right) }-e^{K\left( 2\Delta
-3\right) }}{2\left( 2e^{K}+e^{K\left( 2\Delta -3\right) }+e^{K\left(
1-2\Delta \right) }\right) }.
\end{eqnarray*}%
For two-qubit X state, there are some effective numerical and analytical
expressions\cite{X,QD,X2011}. In this manuscript, we adopt the method in the
Refs.\cite{QD,huang2013} to obtain the QD of the two sites, we don't give
the analytical expression here, because the form is complicated. Then the
numerical result of QD is obtained between two end sites on the fractal
lattices by implementing the decimation RG method\cite{F1,F2,F3}.

Fig. $7$ shows the QD varies the temperature $T$ (with the unit $\left\vert
J\right\vert /k_{\text{B}}$) and anisotropy parameter $\Delta $ in system A
containing $L$ sites. The variation of the QD between two end spins with
temperature $T$ is shown in Figs. $7\left( a\right) $ and $\left( b\right) $%
. It can be seen that the QD exists maximum at $T=0$ and decreases with the
increase of $k_{\text{B}}T/\left\vert J\right\vert $; the QD decreases with
the increase of $L$, and it is almost zero when $L\geq 30$. From Figs. $%
7\left( c\right) $\ and $\left( d\right) $ we can find that\ the QD
decreases with the increasing $\Delta ,$ no matter how large the size of
system is, QD will change to $0$ in the Ising limit $\Delta =1$ and there is
a turning point in the isotropic Heisenberg limit $\Delta =0$. Fig. $8$
shows a continuous change of QD with $T$ and $\Delta $. We can find that in
case of $L=2$, there is a certain mutation in the contour of QD at $\Delta
=0 $; In case of $L>2$, there exists the "cusp" of the contour at $\Delta =0$%
.

Figs. $9$ and $10$ show the QDs vary the temperature $T$ and anisotropy
parameter $\Delta $ in systems B and C. Their basic rule is similar to
system A, but QDs still exist when the size of system is large ($L=2732$ and 
$L=779$\ correspond to systems B and C, respectively). And we find that when 
$T\rightarrow 0$, there is a great difference of QD on both sides of $\Delta
=0$ by the calculation results. When $\Delta >0$ and $T\rightarrow 0$, the
QD is reduced in turn with the increase of $L$ until $L\geq 684$, there
exists almost no QD; When $\Delta \leq 0$ and $T$ $\rightarrow 0$, the QD of 
$L>2$ exists a fixed value, which is shown in Figs. $9\left( a\right) $ and $%
\left( b\right) $. From Figs. $9\left( c\right) $ and $\left( d\right) $, we
can find that there are a turning point of QD for all case at $\Delta =0$.
When $\Delta >0$ and $\Delta <0$, the system tends to be the Ising model\
and XY model, respectively. As shown in Fig. $10$, in system C not only
there is a turning point but also there is a cross point of the QD at $%
\Delta =0$, the turning degree gradually increases with the increase of $L$.
In order to further study the relation of QD with $T$ and $\Delta $, we
prepare the contour plots of QD of systems B (Fig. $11$) and C (Fig. $12$),
which show a continuous change of QD with $\Delta $ and $T$. We can find
that in case of $L=2$, there is a certain mutation in the contour at $\Delta
=0$; In case of $L>2$, there exists the cusp of the contour at $\Delta =0$.

\section{The quantum effect\label{sec5}}

In this section, we discuss the effect of quantum fluctuation and analyze
the error of result which is induced by the noncommutativity. We note that,
we can deal with a classical system straightforwardly, because there will be
no noncommutative. However, for quantum systems, it can not be solved by
such simplifying process because of the noncommutativity between the
Hamiltonians of the neighboring generators\cite{fluctuation}. In the
following, we take system A as an example to discuss this effect which is
the discrepancy existing between the exact result and the approximate one.
Firstly, we assume the generator of system A~(Fig.~$2(a1)$) as the whole
system that we will calculate. In this case, noncommutative is neglected and
the rigorous result can be obtained as $\left( K^{\prime }\left( K,\Delta
\right) ,\Delta ^{\prime }\left( K,\Delta \right) \right) $. For the same
system, we apply the Migdal-Kadanoff method and obtain the approximative
result which is defined as $\left( 2K^{\prime \prime }\left( K,\Delta
;K,\Delta ;K,\Delta \right) ,\Delta ^{^{\prime \prime }}\left( K,\Delta
;K,\Delta ;K,\Delta \right) \right) $. Finally, using the convenient ratios
proposed in Ref.~\cite{important 1985}, we define the errors as%
\begin{align}
E^{K}& \equiv \frac{2K^{\prime \prime }\left( K,\Delta ;K,\Delta ;K,\Delta
\right) }{K^{\prime }\left( K,\Delta \right) }-1,  \label{18} \\
E^{\Delta }& \equiv \frac{\Delta ^{\prime \prime }\left( K,\Delta ;K,\Delta
;K,\Delta \right) }{\Delta ^{\prime }\left( K,\Delta \right) }-1.  \label{19}
\end{align}

The $T$-dependence of $E_{\text{F}}^{K}$,~$E_{\text{F}}^{\Delta }$\ (for the
ferromagnetic case) and $E_{\text{AF}}^{K}$,~$E_{\text{AF}}^{K}\ $(for the
AF case) for typical values of $\Delta $ is presented in Fig.~$13$. As we
can see, both $E^{K}$ and $E^{\Delta }$ tend to zero in the $T\rightarrow
\infty $, and in the Ising limit~($\Delta =1$), both $E^{K}$ and $E^{\Delta
} $ equal zero at all temperature which is in accordance with the previous
conclusion~\cite{Lieb}. In the range of low temperature, the quantum effect
of the AF system is stronger than that of the ferromagnetic system. With the
decrease of $\Delta $, the effect of the fluctuation will be strengthened
for both ferromagnetic and AF systems. The ordering is destroyed by the
quantum fluctuation and this fact is responsible for the reentrant behavior
in the phase diagram of system A. Furthermore, the competition between the
quantum fluctuation and the thermal one is very important. At finite
temperature, quantum fluctuation is usually suppressed in comparison with
thermal one. However, when the temperature is close to zero, the quantum
fluctuation dominate the critical behavior of the system. For systems B and
C, we analyze the phase diagrams and consider that when the N\'{e}el
temperature is relatively high, i.e., the thermal fluctuation is dominated,
the ordering will not be destroyed and there exists phase transition when $%
\Delta $ is close to zero.

\section{Summary}

We have investigated the phase diagrams, critical phenomena and QDs of the
spin-$1/2$ AF Heisenberg model on the DH systems A, B and C. For system A,
we have found the reentrant behavior in the critical line and our results
also show that $T_{\text{N}}\sim \frac{1}{\ln \left( \Delta _{\text{c}%
}-\Delta \right) }$ when $T$ is close to zero. Moreover, it is found that
the N\'{e}el temperature of system B tends to zero and there exists finite
temperature phase transition in system C. We have also found that no matter
how large the size of system is, QD will change to zero in the Ising limit
and there is a turning point of QD in the isotropic Heisenberg limit $\Delta
=0$ where there is a "peak" of the contour for the systems studied.\ And the
QD decreases with the increase of $L$, and it is almost zero in system A
when $L\geq 30$. For system B and C, the QDs still exist when the size of
system is larger than that of system A. In the end, taking system A as an
example, the error of result is analyzed and our result indicate that
quantum fluctuation is the cause of the reentrant behavior in the phase
diagram.

\begin{acknowledgments}
This work is supported by the National Natural Science Foundation of China
under Grants No. 11675090 and No. 11905095; the Shandong Natural Science
Foundation under Grant No. ZR2020MA092. P.-P. Zhang would like to thank
Zhong-Qiang Liu, Yue Li, Zhen-Hui Sun, Li-Zhen Hu and Jing Wang for fruitful
discussions and useful comments.
\end{acknowledgments}

\begin{center}
{\Large Figure Captions}
\end{center}

\bigskip

Fig. 1 The structure of three DH lattices. (a) The dimers. (b) The
generators. (c) The generative lattices. They all start from the dimer\ and
generate the lattices in an iterative manner.

\bigskip

Fig.~2 The procedure of RG transformation. (a) System A: using the
real-space RG method straightforwardly. (b) System B: the first step from
(b1) to (b2) is the equivalent transformation and the second step from (b2)
to (b3) is the RG transformation. \bigskip

\bigskip

Fig.~3 Phase diagram of the AF and ferromagnetic Heisenberg models of system
A. The red line corresponds to the AF Heisenberg case and the inset shows
the phase diagram of ferromagnetic case. O(D) stands for ordered
(disordered) phase. The open circle and the full square denote the unstable
and fully stable fixed points, respectively.

\bigskip

Fig.~4 The scaling behaviors of system A when $T\rightarrow 0$. (a) and (b)
represent the critical temperature $T_{\text{c}}$ of ferromagnetic case and
the N\'{e}el temperature $T_{\text{N}}$\ of AF case vary with $\Delta $,
respectively.

\bigskip

Fig.~5 Phase diagrams of the three systems. There exists a reentrant
behavior in system A, the N\'{e}el temperature tends to zero as decreasing $%
\Delta $ in system B and there exists finite temperature phase transition in
the isotropic Heisenberg limit $\Delta =0$ in system C.

\bigskip

Fig.~6 The scaling behavior of system\ B in the $\Delta \rightarrow 0$
limit. The N\'{e}el temperature $T_{\text{N}}$ is inversely proportional to $%
\ln \Delta $.

\bigskip

Fig. 7 The QD between two end spins on system A with $L$ sites varies with $%
T $ (with the unit $\left\vert J\right\vert /k_{\text{B}}$, the same below)
and anisotropy parameter $\Delta $. (a) and (b) the QD varies with $T$, (c)
and (d) the QD varies with $\Delta $. The QD for different $L$ cases has a
turning point at $\Delta =0$.

\bigskip

Fig. 8 The contour plot of QD of system A with $L$ sites varies with $T$ and 
$\Delta $. When $L=2$, there is a certain mutation in the contour of QD at $%
\Delta =0$; When $L>2$, there exists the "cusp" in the contour of QD at $%
\Delta =0$.

\bigskip

Fig. 9 The QD between two end spins of system B with $L$ sites varies with
with $T$ and $\Delta $. (a) and (b) the QD varies with $T$, (c) and (d) the
QD varies with $\Delta $. The QD for different $L$ cases has a turning point
at $\Delta =0$.

\bigskip

Fig. 10 The QD of system C varies with $T$ and $\Delta $. (a) and (b) the QD
varies with $T$, (c) and (d) the QD varies with $\Delta $. The QD for
different $L$ cases has a cross point and a turning point at $\Delta =0$.

\bigskip

Fig. 11 The contour plot of system B with $L$ sites varies with $T$ and $%
\Delta $. When $L=2$, there are a mutation in the contour of QD at $\Delta
=0 $; When $L>2$, there exists the "cusp" in the contour of QD at $\Delta =0$%
.

\bigskip

Fig. 12 The contour plot of system C. When $L=2$, there is a mutation in the
contour of QD at $\Delta =0$; When $L>2$, there exists the "cusp" in the
contour of QD at $\Delta =0$.

\bigskip

Fig.~13 Thermal dependence of the errors $E^{K}$ and $E^{\Delta }$ for the
ferromagnetic and AF systems defined by Eq.~(\ref{18}) and Eq.~(\ref{19}),
for typical values of $\Delta $. (a) and (b) correspond to $E^{K}$, (c) and
(d) scorrespond to $E^{\Delta }$. It can be seen obviously that $E^{K}$ and $%
E^{\Delta }$ for the ferromagnetic system are larger than the AF system at
low temperature. Our calculations also show $E^{K}$ and $E^{\Delta }$ equal
zero in the Ising limit~($\Delta =1$).

\bigskip

\bigskip

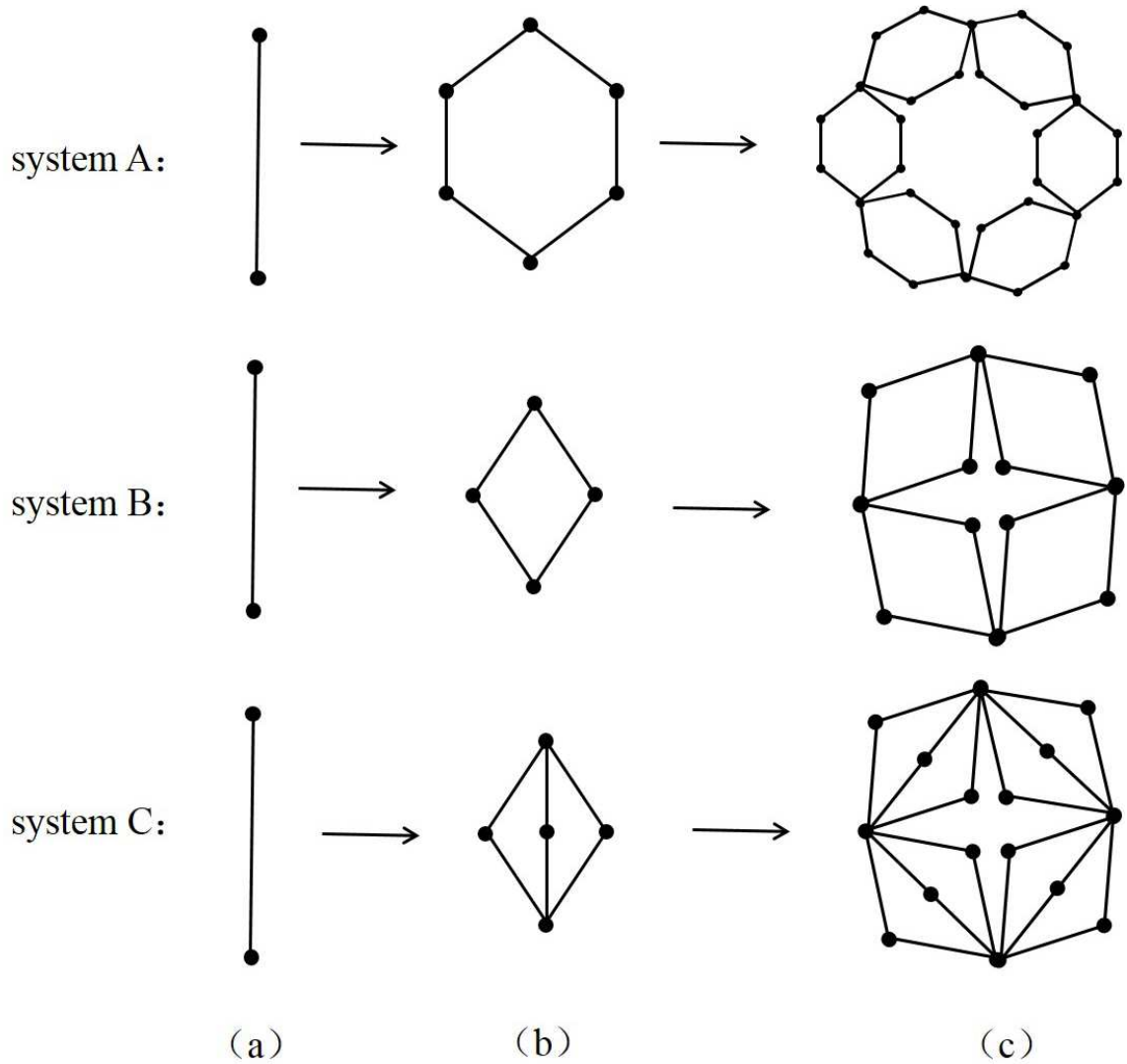
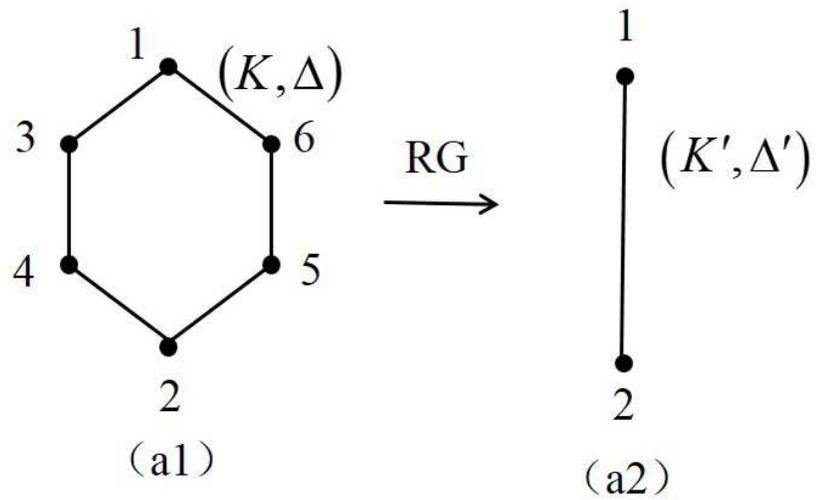
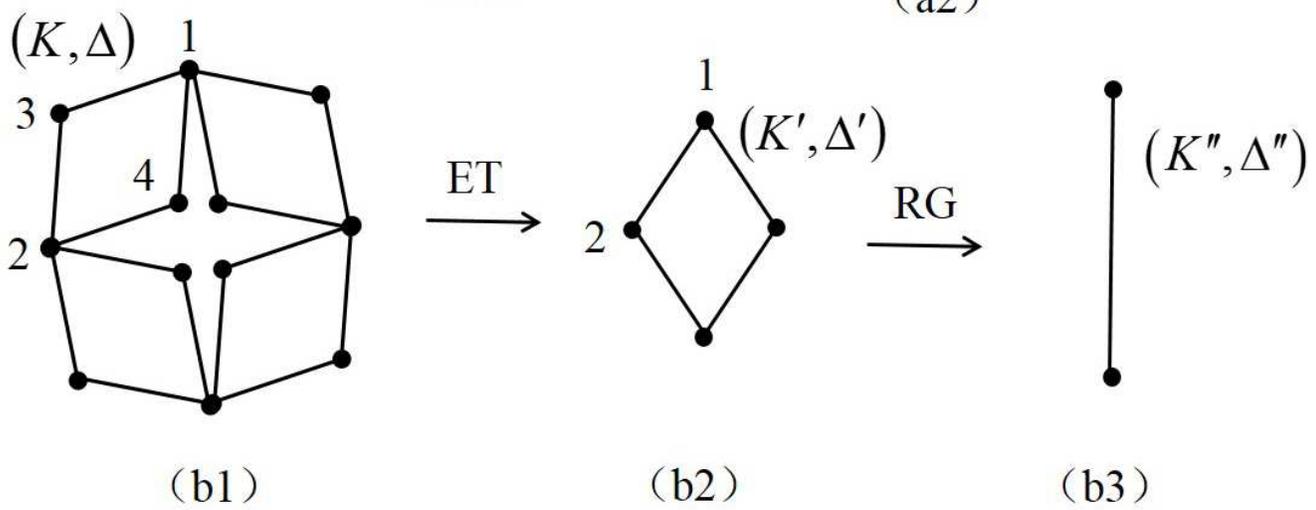
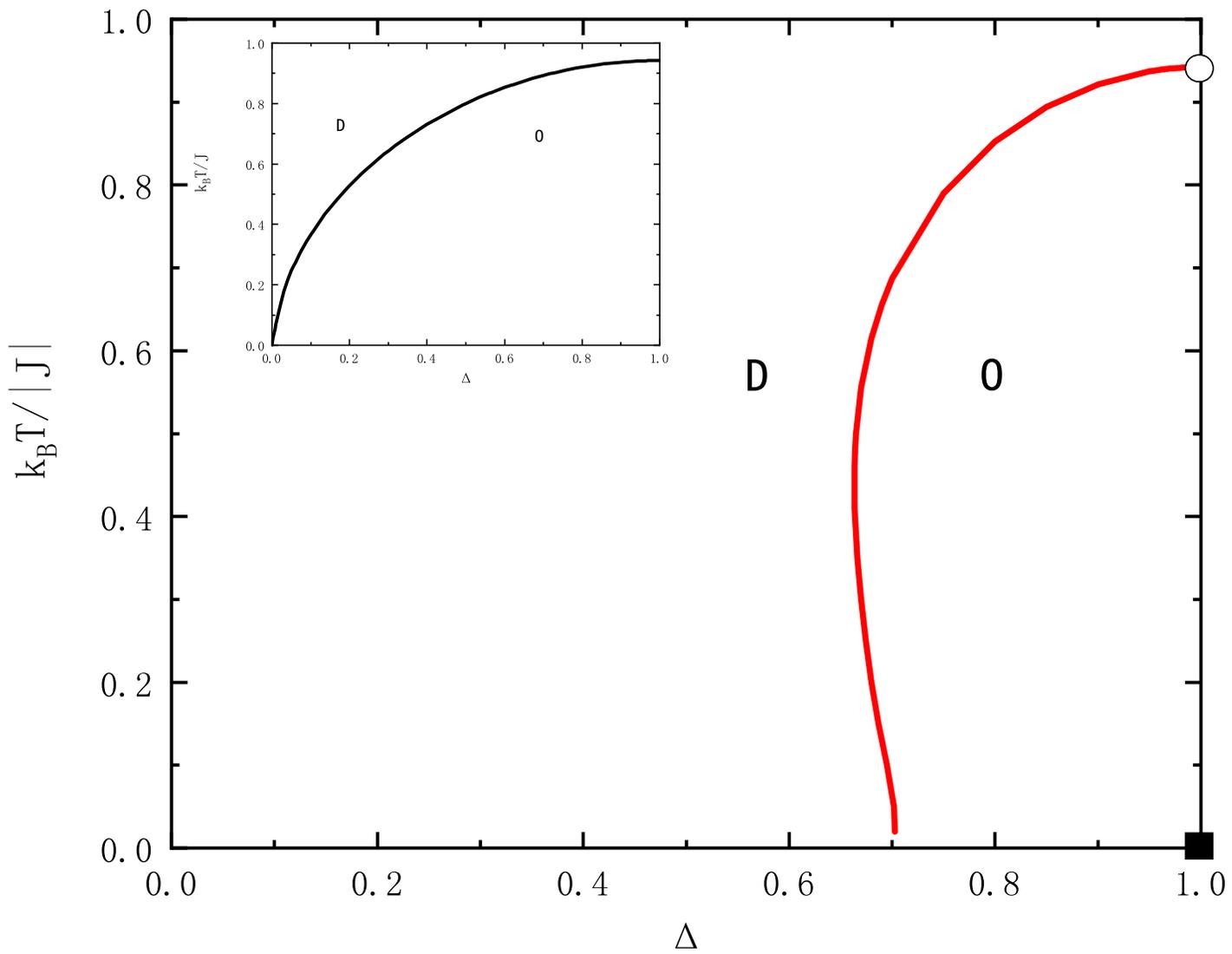
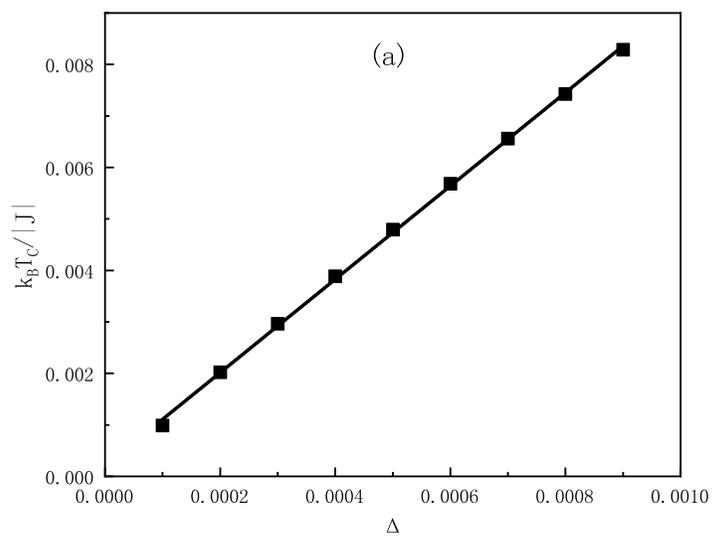
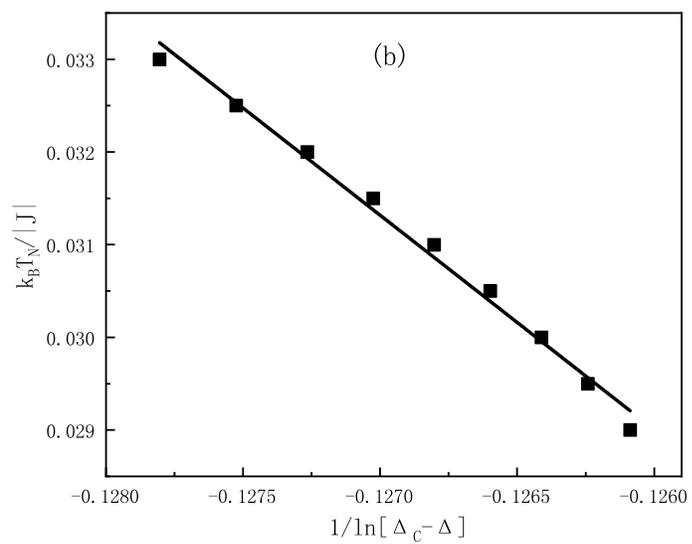
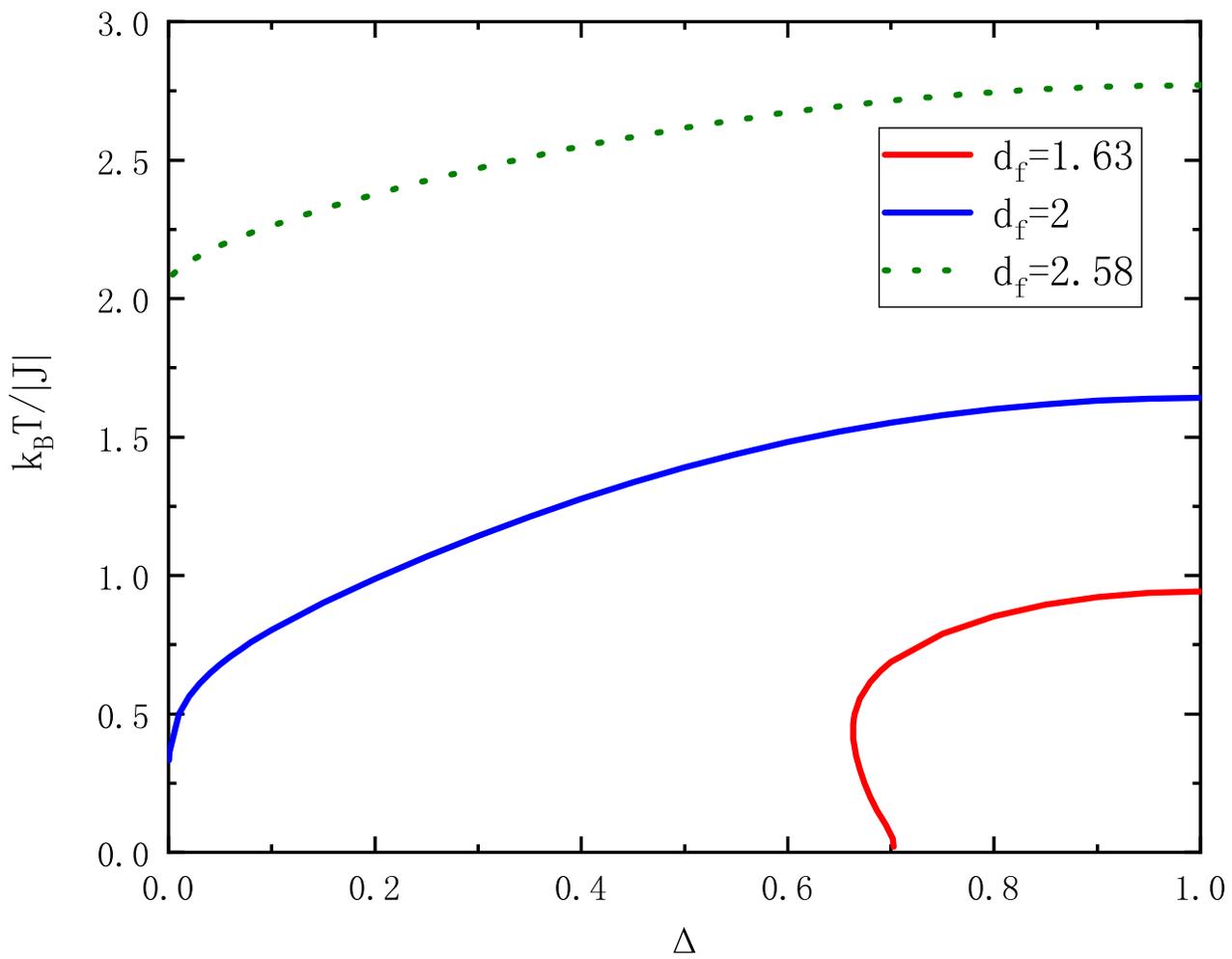
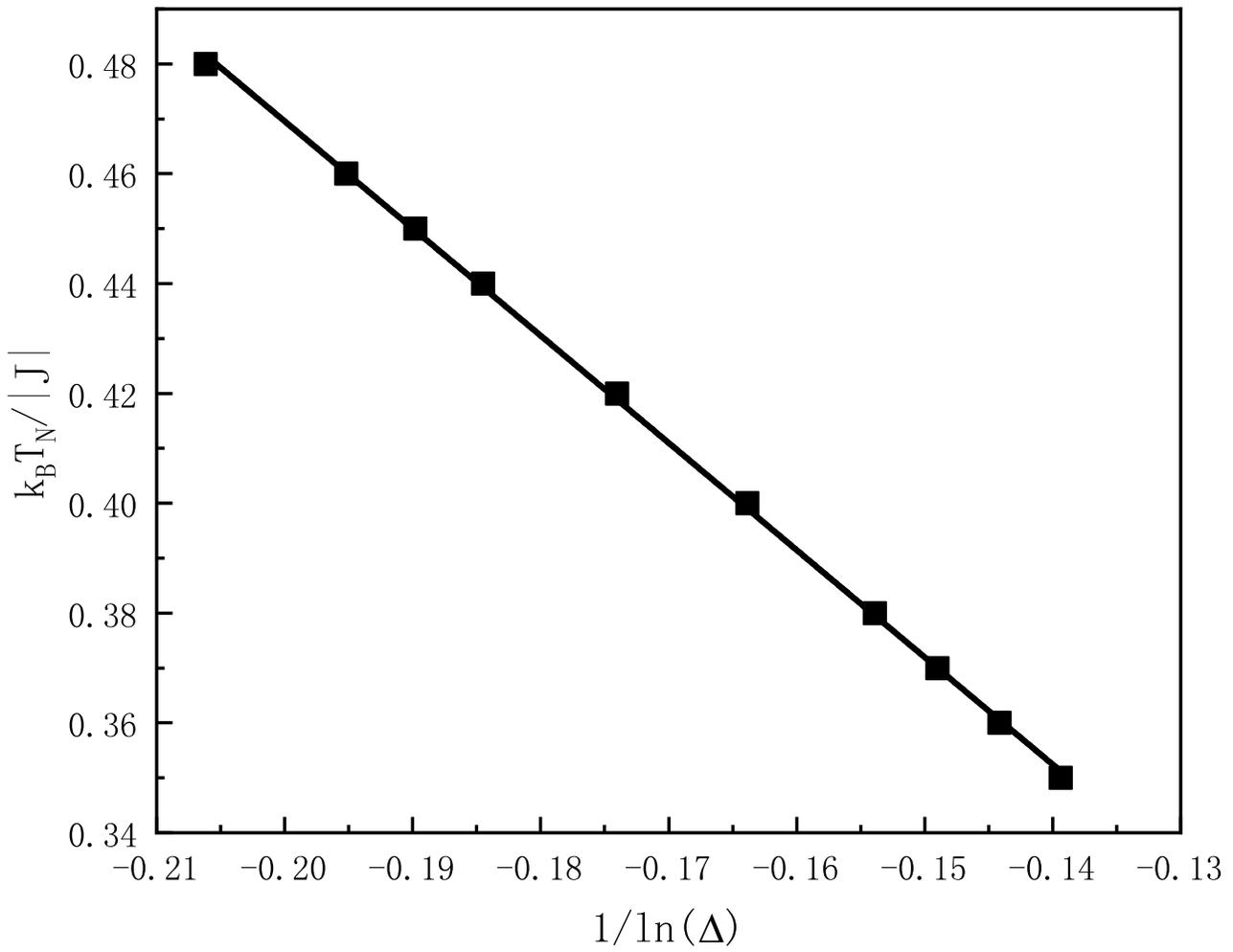
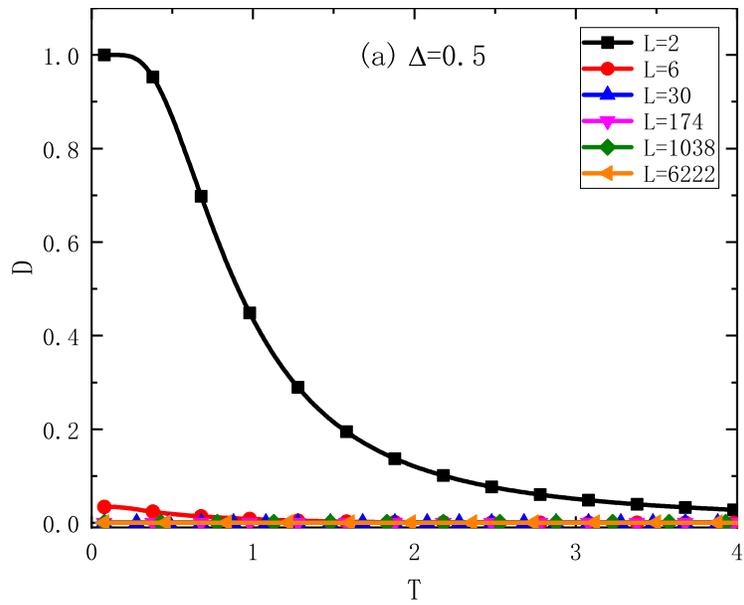
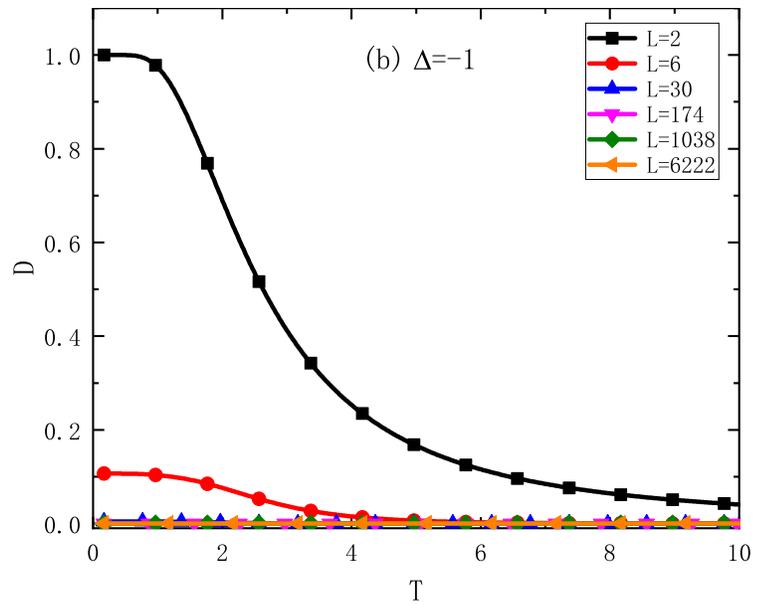
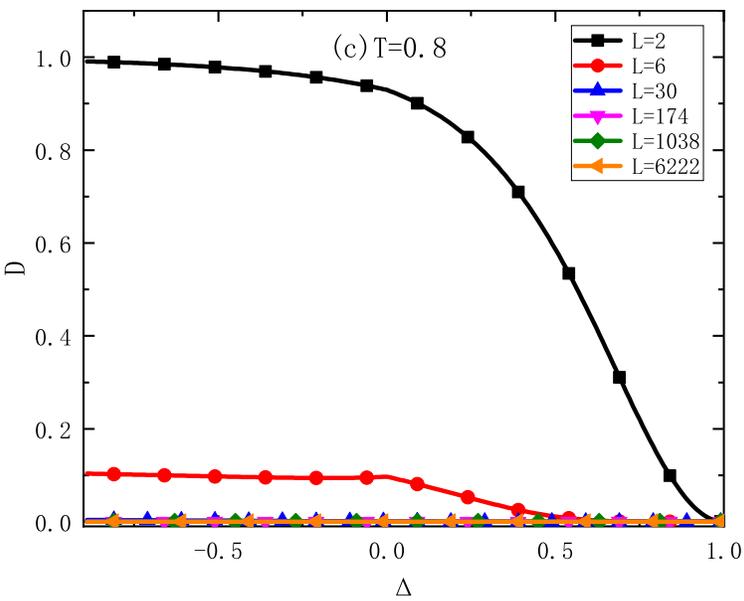
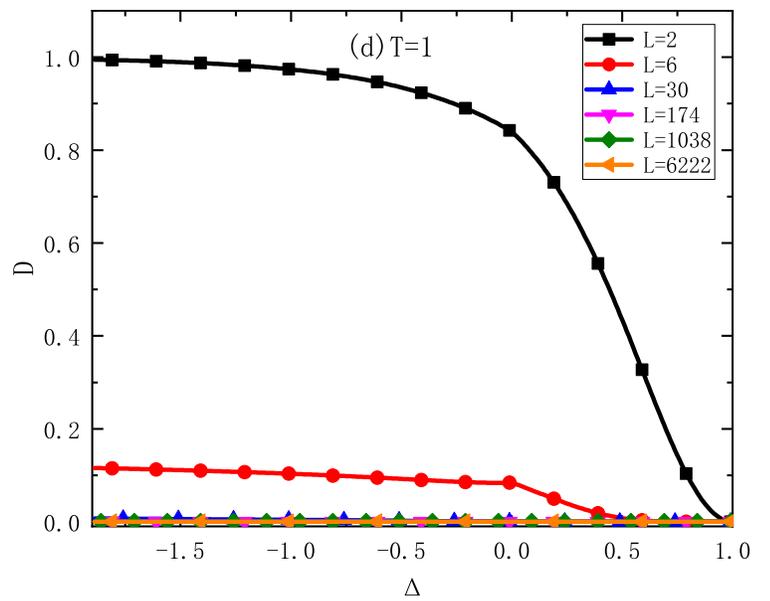
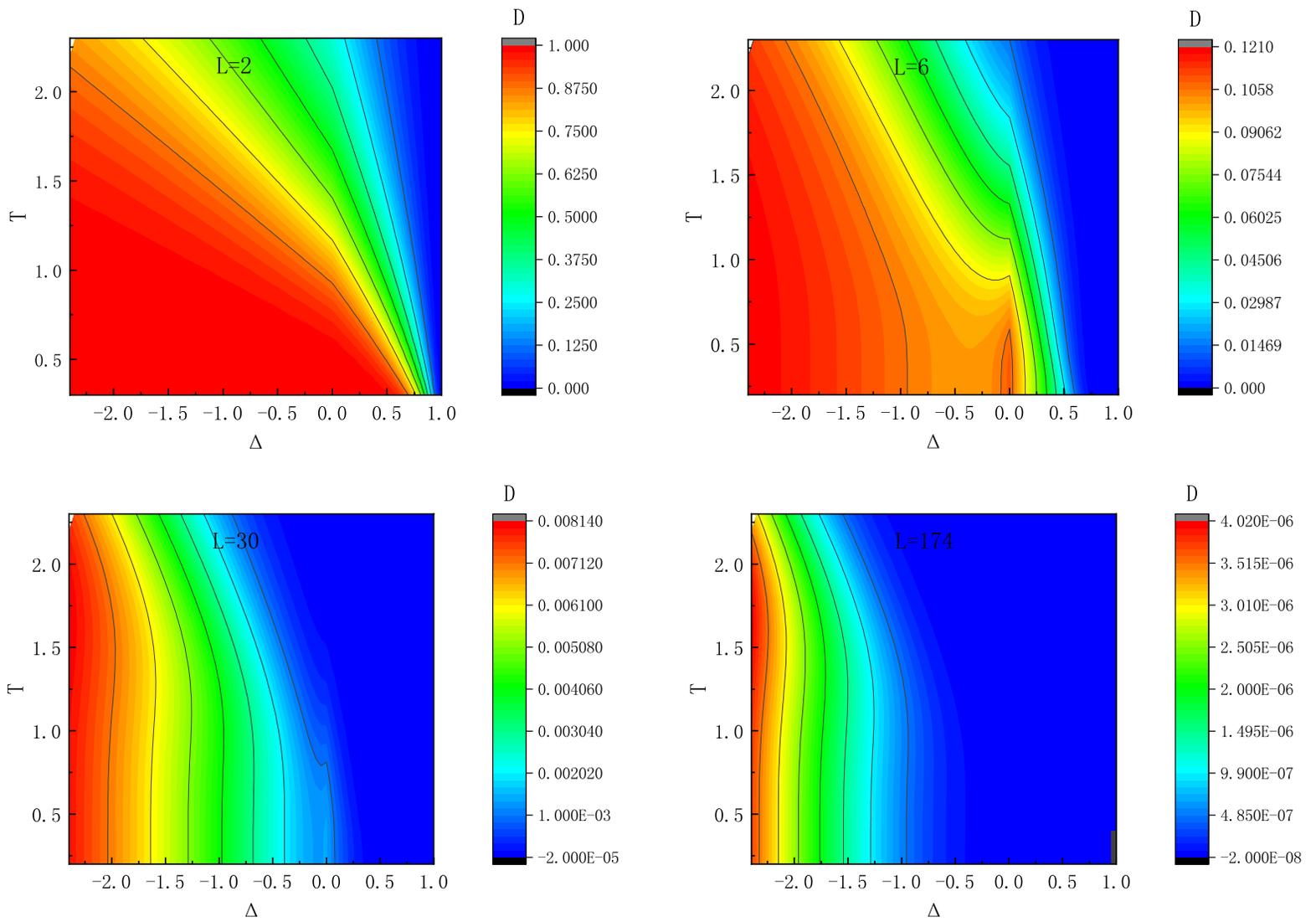
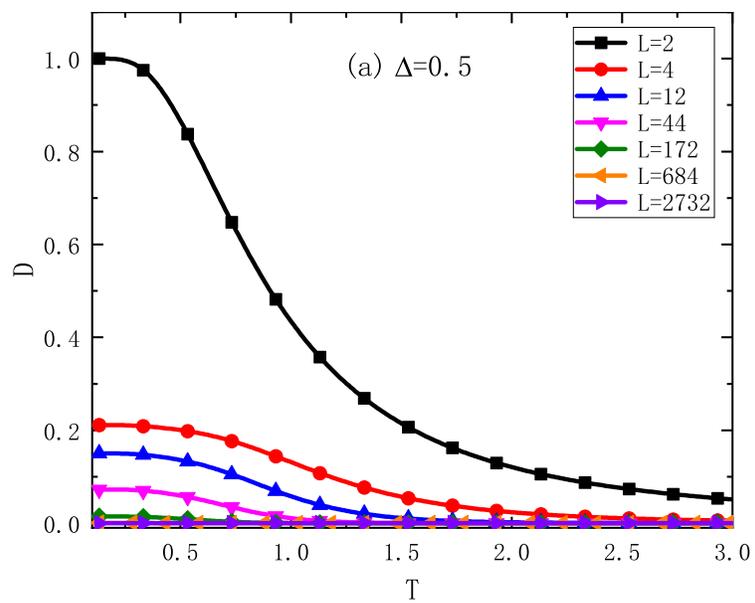
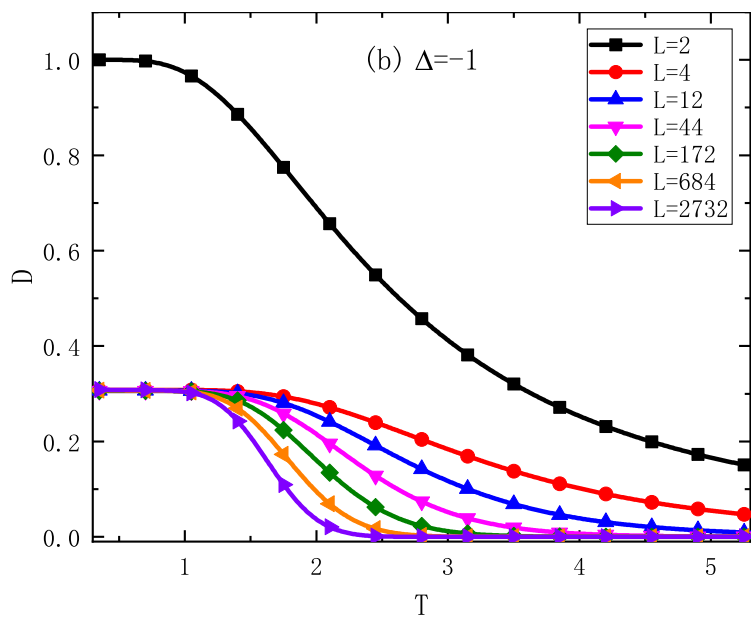
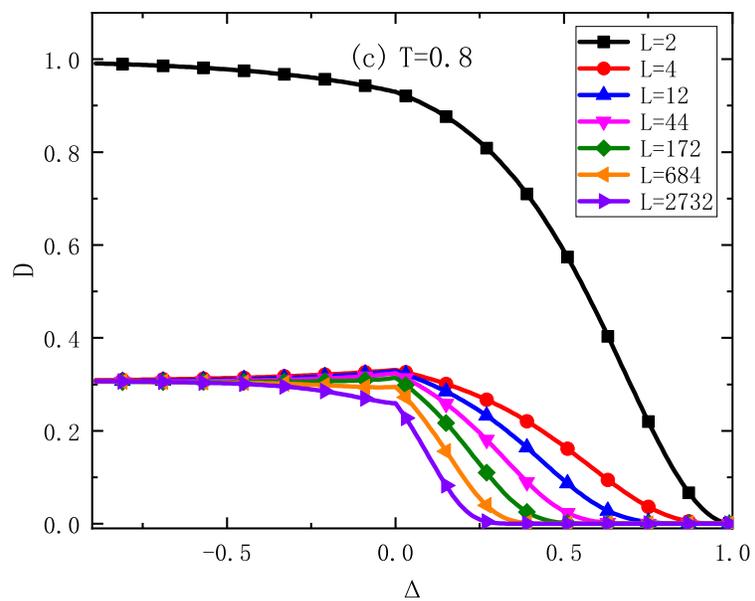
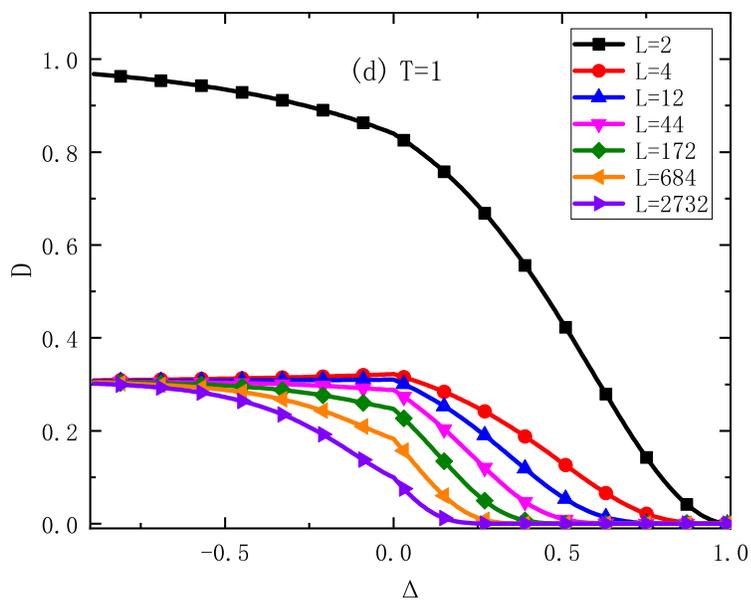
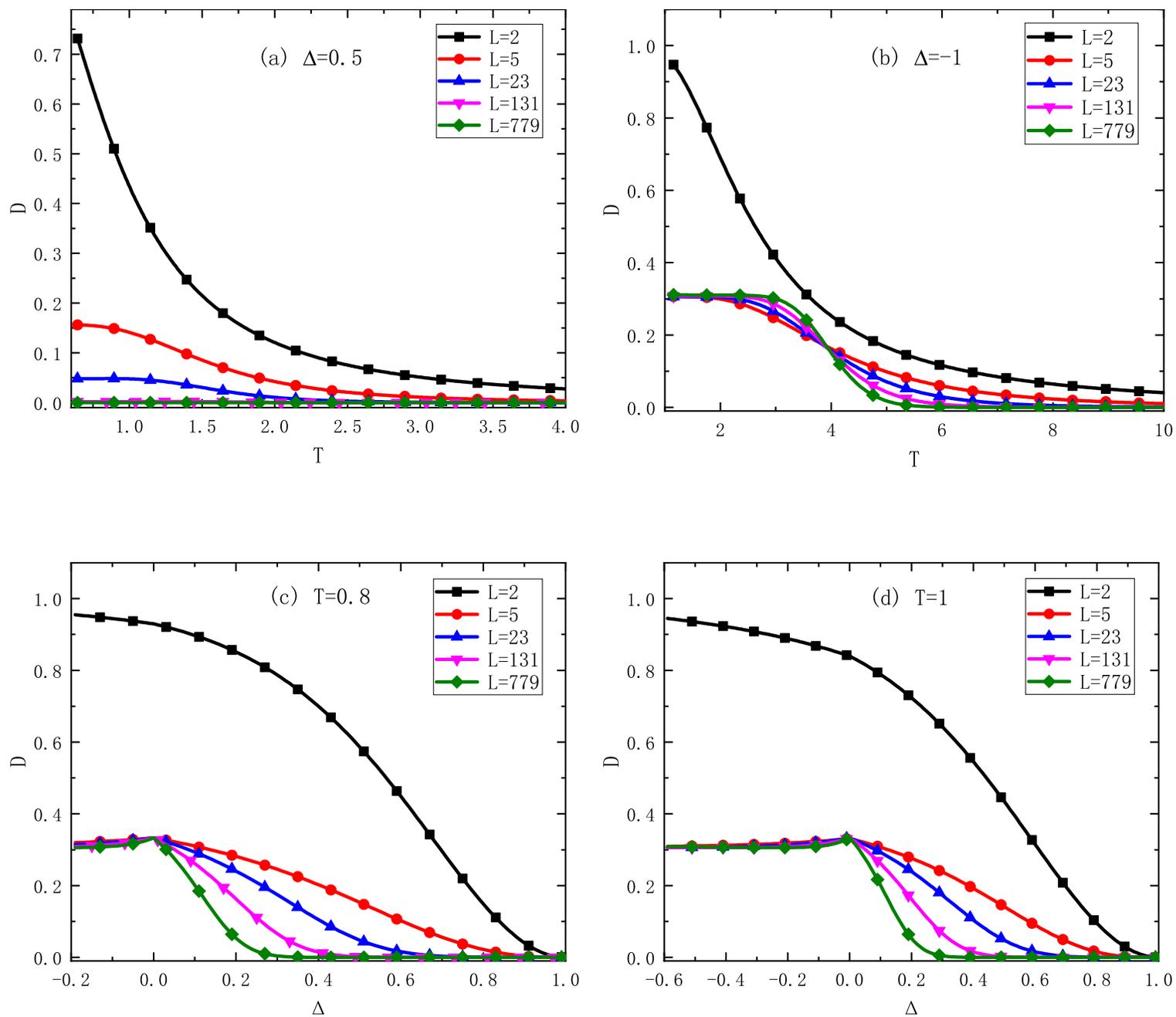
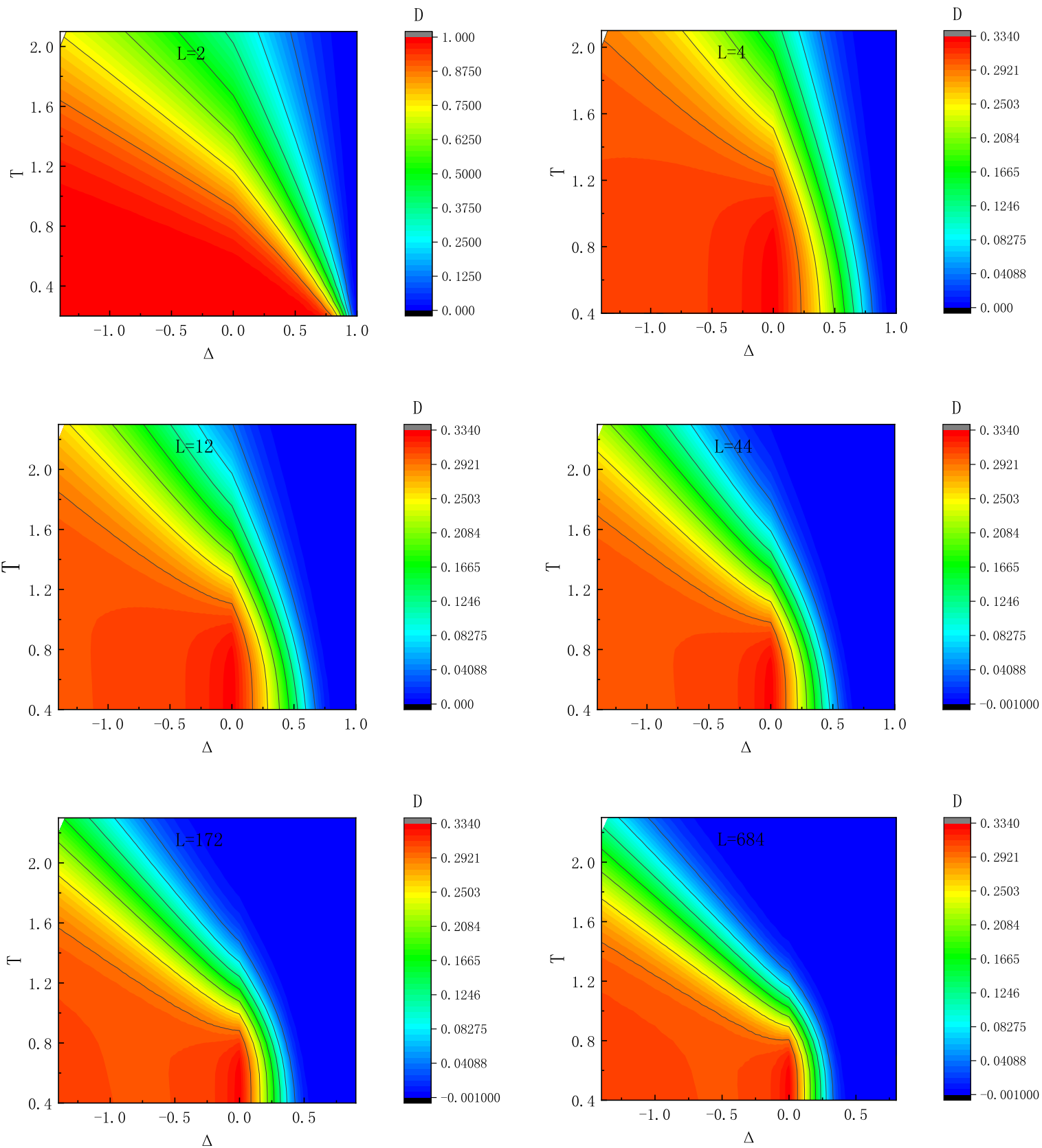
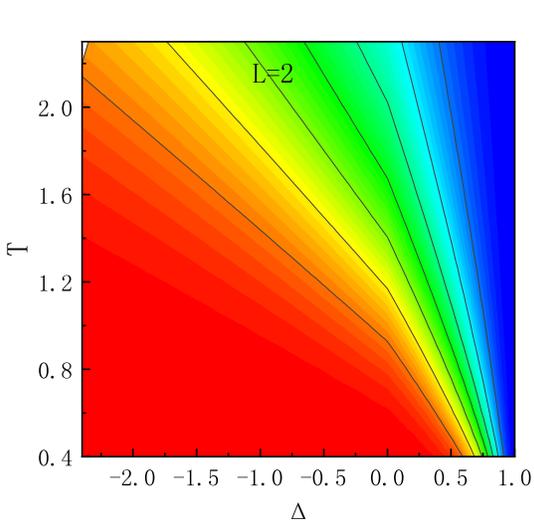
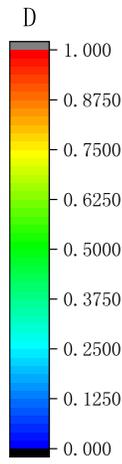
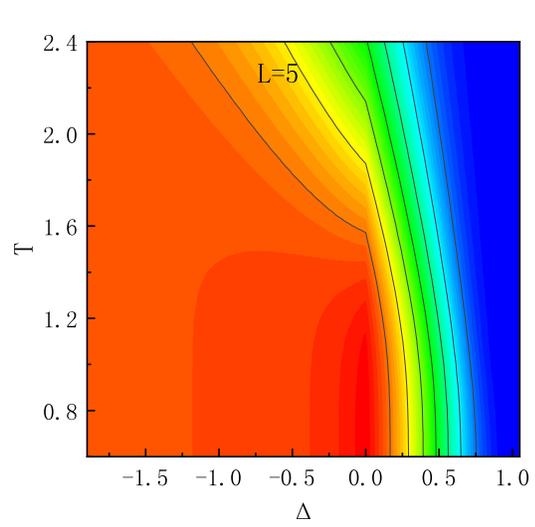
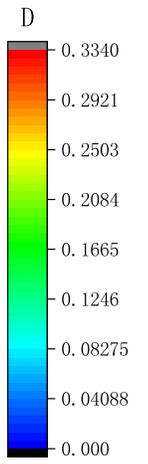
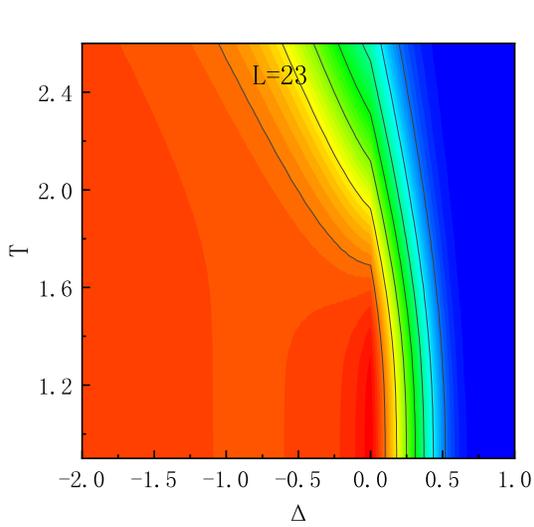
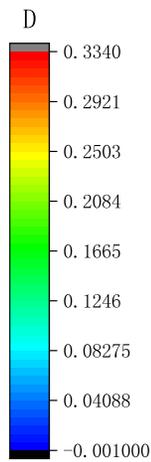
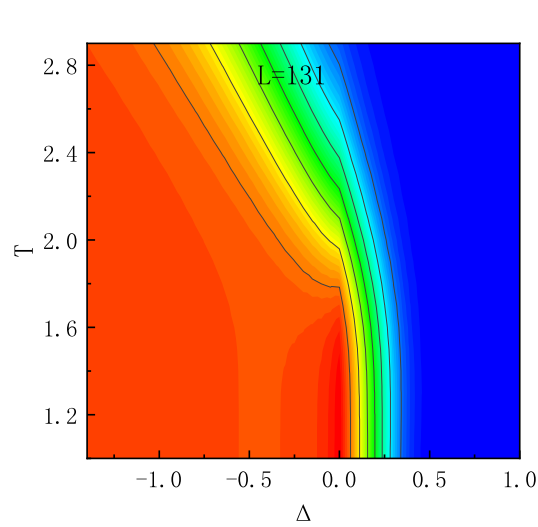
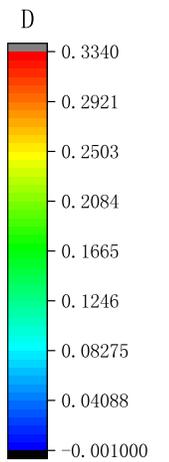
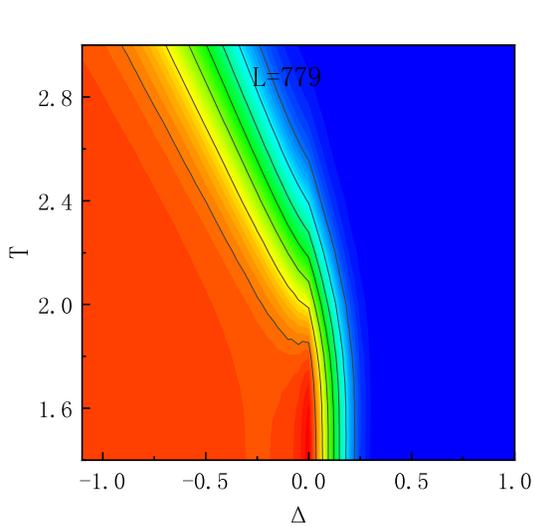
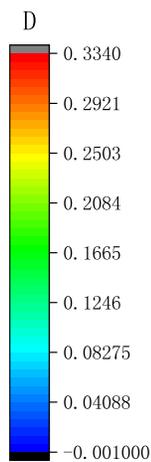
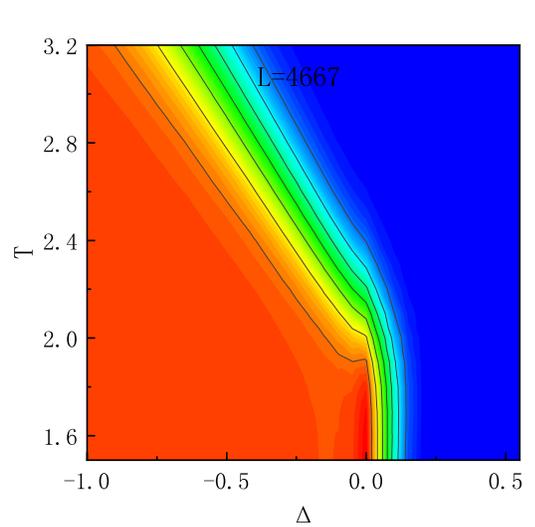
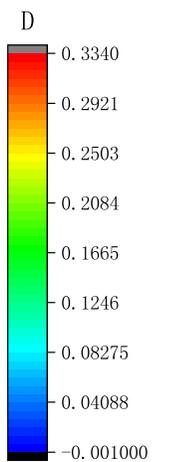
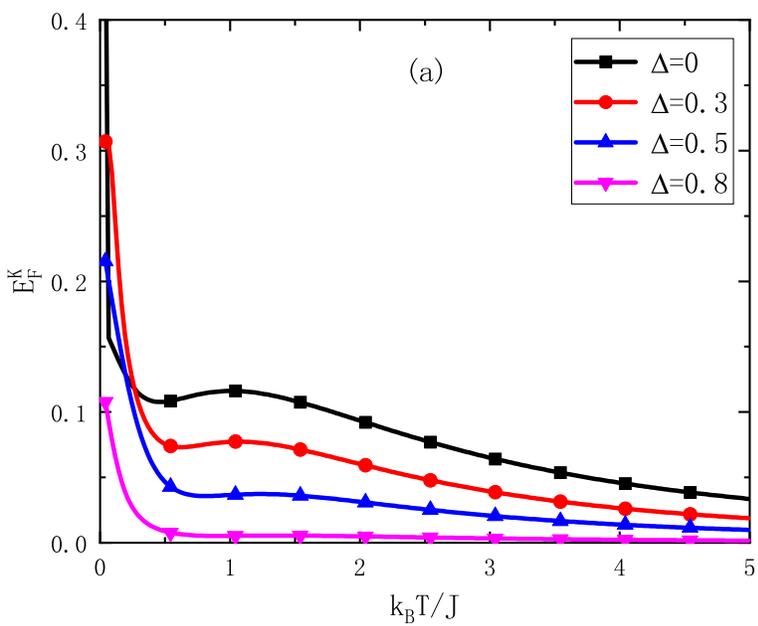
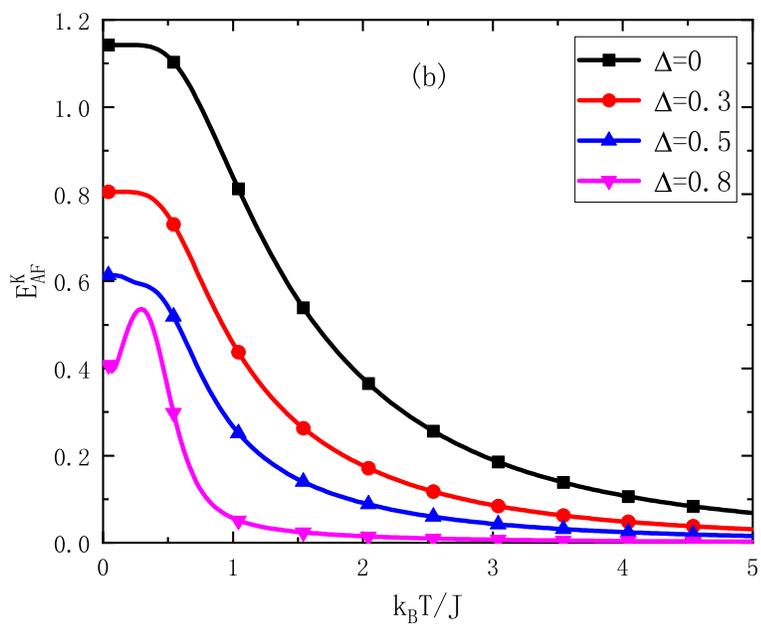
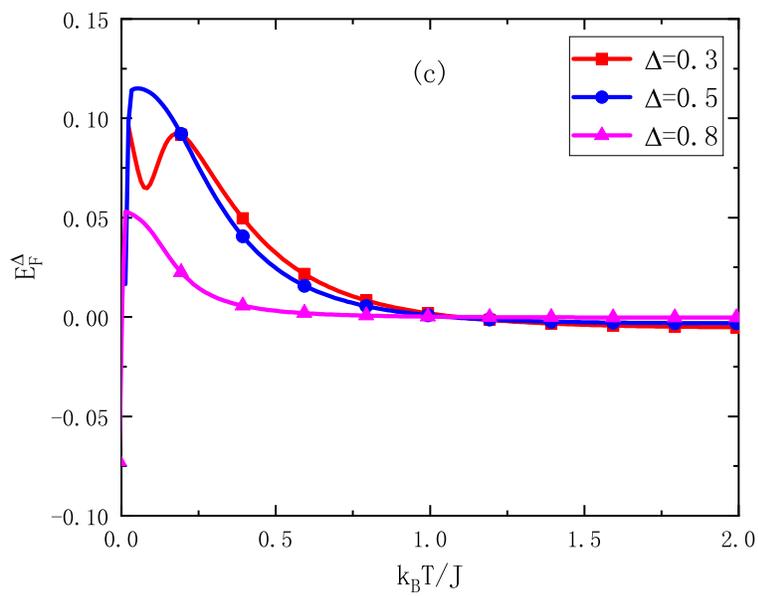
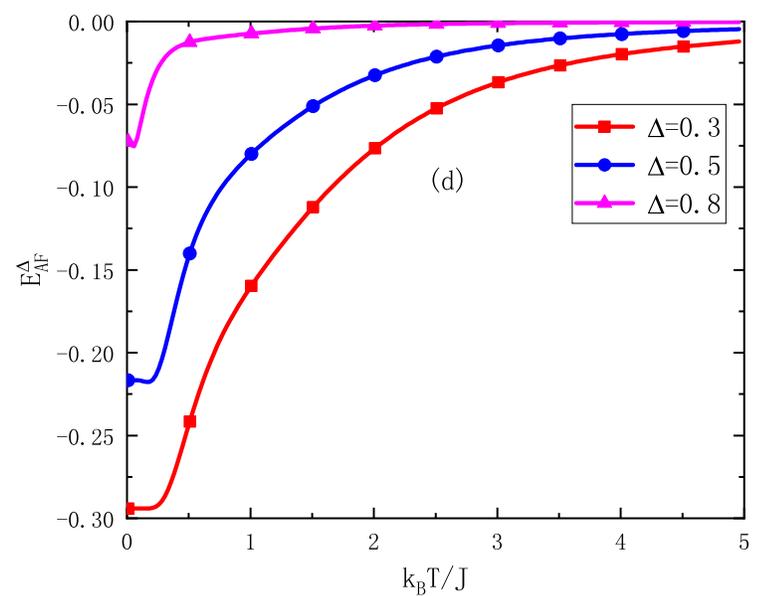
\end{document}